\documentclass[10pt,twocolumn,twoside]{IEEEtran}
\usepackage{amssymb}
\usepackage{amsfonts}
\usepackage{amsmath}
\usepackage{cite}
\usepackage[dvips]{graphicx}
\usepackage{color}
\usepackage{comment}
\usepackage{makecell}

\begin{document}
% ===only use in IEEE format : BEGIN ====
\newtheorem{definition}{\it Definition}%[section]
\newtheorem{theorem}{\bf Theorem}%[section]
\newtheorem{lemma}{\it Lemma}
\newtheorem{corollary}{\it Corollary}
\newtheorem{remark}{\it Remark}
\newtheorem{example}{\it Example}
\newtheorem{case}{\bf Case Study}
\newtheorem{assumption}{\it Assumption}
\newtheorem{property}{\it Property}
\newtheorem{proposition}{\it Proposition}
% ===only use in IEEE format : END ====

\newcommand{\hP}[1]{{\boldsymbol h}_{{#1}{\bullet}}}
\newcommand{\hS}[1]{{\boldsymbol h}_{{\bullet}{#1}}}

\newcommand{\ba}{\boldsymbol{a}}
\newcommand{\baq}{\overline{q}}
\newcommand{\bA}{\boldsymbol{A}}
\newcommand{\bb}{\boldsymbol{b}}
\newcommand{\bB}{\boldsymbol{B}}
\newcommand{\bc}{\boldsymbol{c}}
\newcommand{\bcO}{\boldsymbol{\cal O}}
\newcommand{\bh}{\boldsymbol{h}}
\newcommand{\bH}{\boldsymbol{H}}
\newcommand{\bl}{\boldsymbol{l}}
\newcommand{\bm}{\boldsymbol{m}}
\newcommand{\bn}{\boldsymbol{n}}
\newcommand{\bo}{\boldsymbol{o}}
\newcommand{\bO}{\boldsymbol{O}}
\newcommand{\bp}{\boldsymbol{p}}
\newcommand{\bq}{\boldsymbol{q}}
\newcommand{\bR}{\boldsymbol{R}}
\newcommand{\bs}{\boldsymbol{s}}
\newcommand{\bS}{\boldsymbol{S}}
\newcommand{\bT}{\boldsymbol{T}}
\newcommand{\bw}{\boldsymbol{w}}

\newcommand{\balpha}{\boldsymbol{\alpha}}
\newcommand{\bbeta}{\boldsymbol{\beta}}
\newcommand{\bOmega}{\boldsymbol{\Omega}}
\newcommand{\bTheta}{\boldsymbol{\Theta}}
\newcommand{\bphi}{\boldsymbol{\phi}}
\newcommand{\btheta}{\boldsymbol{\theta}}
\newcommand{\bvarpi}{\boldsymbol{\varpi}}
\newcommand{\bpi}{\boldsymbol{\pi}}
\newcommand{\bpsi}{\boldsymbol{\psi}}
\newcommand{\bxi}{\boldsymbol{\xi}}
\newcommand{\bx}{\boldsymbol{x}}
\newcommand{\by}{\boldsymbol{y}}

\newcommand{\cA}{{\cal A}}
\newcommand{\bcA}{\boldsymbol{\cal A}}
\newcommand{\cB}{{\cal B}}
\newcommand{\cE}{{\cal E}}
\newcommand{\cG}{{\cal G}}
\newcommand{\cH}{{\cal H}}
\newcommand{\bcH}{\boldsymbol {\cal H}}
\newcommand{\cK}{{\cal K}}
\newcommand{\cO}{{\cal O}}
\newcommand{\cR}{{\cal R}}
\newcommand{\cS}{{\cal S}}
\newcommand{\dcS}{\ddot{{\cal S}}}
\newcommand{\ds}{\ddot{{s}}}
\newcommand{\cT}{{\cal T}}
\newcommand{\cU}{{\cal U}}
\newcommand{\wt}[1]{\widetilde{#1}}

\newcommand{\mA}{\mathbb{A}}
\newcommand{\mE}{\mathbb{E}}
\newcommand{\mG}{\mathbb{G}}
\newcommand{\mR}{\mathbb{R}}
\newcommand{\mS}{\mathbb{S}}
\newcommand{\mU}{\mathbb{U}}
\newcommand{\mV}{\mathbb{V}}
\newcommand{\mW}{\mathbb{W}}

\newcommand{\uq}{\underline{q}}
\newcommand{\ubq}{\underline{\boldsymbol q}}

\newcommand{\red}[1]{\textcolor[rgb]{1,0,0}{#1}}
\newcommand{\gre}[1]{\textcolor[rgb]{0,1,0}{#1}}
\newcommand{\blu}[1]{\textcolor[rgb]{0,0,0}{#1}}

% paper title
% can use linebreaks \\ within to get better formatting as desired
\title{Towards Self-learning Edge Intelligence in 6G} % Integration of Artificial Intelligence into 6G with Federated Learning} % over Licensed and Unlicensed Bands}

\author{Yong~Xiao, %\IEEEmembership{Senior~Member,~IEEE},
Guangming~Shi, Yingyu Li, Walid Saad, and H. Vincent Poor %\IEEEmembership{Senior~Member,~IEEE},
%Marwan~Krunz, %\IEEEmembership{Fellow, IEEE}, %and Ross~Murch, \IEEEmembership{Fellow, IEEE}

\thanks{Y. Xiao and Yingyu Li (corresponding author) are with the School of Electronic Information and Communications at the Huazhong University of Science and Technology, Wuhan, China 430074 (e-mail: \{yongxiao, liyingyu\}@hust.edu.cn).

G. Shi is with the School of Artificial Intelligence, the Xidian University, Xi'an, Shaanxi 710071, China (e-mail: gmshi@xidian.edu.cn),

%M. Krunz is with the Department of Electrical and Computer Engineering at the University of Arizona, Tucson, AZ 85710 (e-mail: krunz@email.arizona.edu),

%K-C Chen is with the Department of Electrical Engineering, University of South Florida, Tampa, FL 33620 (e-mail: kwangcheng@usf.edu),

W. Saad is with the Department of Electrical and Computer Engineering at Virginia Tech, Blacksburg, VA 24061 (e-mail: walids@vt.edu). %,

H. V. Poor is with the School of Engineering and Applied Science, Princeton University, Princeton, NJ 08544 (e-mail: poor@princeton.edu).
%R. Murch is with the Department of Electronic and Computer Engineering, the Hong Kong University of Science and Technology, Hong Kong (e-mail: eermurch@ust.hk).
}
}

\maketitle

\begin{abstract}
Edge intelligence, also called edge-native artificial intelligence (AI), is an emerging technological framework focusing on seamless integration of AI, communication networks, and mobile edge computing. It has been considered to be one of the key missing components in the existing 5G network and is widely recognized to be one of the most sought-after functions for tomorrow's wireless 6G cellular systems. In this article, we identify the key requirements and challenges of edge-native AI in 6G. A self-learning architecture based on self-supervised Generative Adversarial Nets (GANs) is introduced to \blu{demonstrate the potential performance improvement that can be achieved by automatic data learning and synthesizing at the edge of the network}. We evaluate the performance of our proposed self-learning architecture in a university campus shuttle system connected via a 5G network. Our result shows that the proposed architecture has the potential to identify and classify unknown services that emerge in edge computing networks. Future trends and key research problems for self-learning-enabled 6G edge intelligence are also discussed. %as a major enabler of 6G to unleash the full potential of network intelligentization.
 %A self learning-based architecture  a possible research roadmap for 6G edge intelligence, from the perspective of self-learning AI. Potential requirements and challenges of edge-native AI in 6G have been identified.
 %The potential of self-learning AI to address some other challenging issues of 6G edge intelligence is also discussed.

\end{abstract}
%\vspace{-0.2in}

\section{Introduction}
\label{Section_Introduction}

The wireless networking landscape is witnessing an unprecedented evolution that has led to the deployment of the fifth generation (5G), the latest iteration of mobile technology that promises to support a plethora of innovative services including the Internet-of-Things (IoT), autonomous vehicles, Augmented Reality/Virtual Reality (AR/VR), among others. Simultaneously, a broad range of research is being initiated to look into the sixth generation (6G) of wireless cellular systems \cite{Saad20196GSurvey} whose primary goals include not only a much improved data transportation network but also a highly intelligent and fully autonomous human-oriented system.  %providing higher spectrum efficiency and reliable connectivity for the next generation of connected autonomy applications.
In particular, 6G will revolve around a new vision of {\em ubiquitous AI}, a hyper-flexible architecture that brings human-like intelligence into every aspect of networking systems.

%The applications of AI in wireless systems have attracted significant interest recently due to the remarkable success of AI in a range of real-world tasks such as computer-based vision and Natural language processing. %Compared to the existing cloud-based AI such as the mobile voice-assistance that has already been widely available in 4G networks, the research communities are now actively exploring potential AI-inspired functional modules and applications in 5G to improve the performance and efficiency of the networking architecture.
There are already initiatives to promote the applications of AI in 5G. In particular, ITU-T has established the focus group (FG) on `machine learning for future networks including 5G' (ML5G)\cite{ITUFGML5G} to promote unified architectural development and interface design for the cost-effective integration of machine learning (ML) into 5G and future networks. 3GPP is also reported to work on new AI-inspired functional modules to monitor and improve the performance of the service-based architecture (SBA).

Despite its great potential, we have yet to observe a wide spread deployment of AI in wireless systems due to the following three major challenges:
\begin{itemize}
\item[(1)] {\bf Limited Resources:} The storage space and computational resources needed for executing AI algorithms often exceed those of existing wireless systems. %As the volume of data and wireless service requirements continue to grow in an unprecedented speed, the gap between the required and available storage and processing resources for next generation wireless networking systems will become even larger.
\item[(2)] {\bf Lack of High-quality Labeled Data:} Most existing AI algorithms require a large number of (high-quality) labeled data for learning and model training. %which is not always available in wireless systems
%Although the volume of data generated by wireless networks is expected to reach massive-scale, most of these data is unlabeled raw data.
Meanwhile, the massive volume of data generated by wireless networks is mostly unlabeled raw data. Manually labeling these datasets is time-consuming and, in most cases, impractical. This challenge is further exacerbated by the fact that wireless network data is highly random in nature and the required quantity and quality of the labeled data for model training and construction are closely related to a range of uncontrollable and unpredictable factors such as geographic locations, operating frequency, distribution of network infrastructure, user mobility, software and hardware configurations, and others.
\item[(3)] {\bf Lack of AI Optimized Architecture:} The existing wireless network architecture has not been originally designed with AI-inspired applications and services in mind. Deploying resource-consuming AI solutions may strain the capacity of the wireless infrastructure, which has already been overloaded with resource-hungry applications. Currently, there is still a lack of an AI-native networking architecture that can strike a  balance  between the resource need for delivering AI functions and that for supporting the fast-growing number of mobile applications with stringent requirements.
\end{itemize}

%a number of reasons. First, the storage space and computational capacities needed for executing AI algorithms often exceed those of existing wireless network infrastructure. As the volume of data and wireless service requirements continue to grow in an unprecedented speed, the gap between the required and available storage and processing resources for next generation wireless networking systems will become even larger. Second, the wireless network architecture has not been originally designed or optimized with AI-inspired applications and services in mind. Finally,

% has been hindered by several challenges.
%
%Currently, the AI-based application is still in its infancy. However, it is commonly believed that
%
%First,  %Collecting sufficient amount of data in every possible condition is generally impossible.
%Second, % generated by modern mobile networks is massive and
%\begin{itemize}
%\item[] {\bf Storage and Processing Capacity}:
 %Transporting all these data into a location, cloud data center, for a centralized process will not only cause intolerable delay but also has the potential to congest the entire networks.

%\item[] {\bf Architecture is not AI-optimized}:

%
%\item[] {\bf Labeled Dataset}:

%\end{itemize}

One possible solution to address the above challenges, is {\em edge intelligence}, also referred to as edge-native AI\cite{Park2019EdgeIntelligence,Zhu2020EdgeIntelligence}, a novel technological framework focusing on seamless integration of  AI, communication networks, and mobile edge computing\footnote{There are other concepts such as fog computing and multi-access edge computing that convey a similar meaning. % and only subtle difference. %Subtle difference may exist among these terms depending on the specific situation and context.
In this article, we use these terms interchangeably to mean any capable devices other than cloud data centers.}. In particular, by deploying a massive scale of decentralized mobile edge servers to perform AI-based processing and decision making closer to where the data and service requests are generated, edge intelligence lays the foundation for ubiquitous and accelerated AI integration  in the next generation wireless system. %,  %achieve the balance the resource utilization,
%significantly improved service responsiveness and resource utilization, as well as reduced deployment costs and potential security risks.

Edge intelligence is commonly considered to be one of the key missing components in 5G\cite{Zhou2019EdgeIntelligence} and %. With the mobile networking technology being gradually revolve into the vision of ``edge-native AI"-based 6G, and
is well recognized as a major enabler for 6G to unleash the full potential of %``edge-native AI" and bring
network intelligentization\cite{Chen2019MLforWireless, XY2018TactileInternet}. %, catalyzing a constellation of innovative solutions for driving human society toward a more sustainable and prosperous future.
The role of AI-enabled applications in terms of architecture, functional components, requirements, and applications %suggested by ITU and some recent industrial and academia white papers\cite{ITU2019Network2030, Dang20206GSurvey, Letaief2019Roadmapto6G, Tariq20196GSurvey, Saad20196GSurvey}
in 6G\cite{Letaief2019Roadmapto6G} compared with those in 4G and 5G are summarized in Table \ref{Tabel_5G6GCompare}. %The roadmap for 6G evolution according to ITU are described in Figure \ref{Figure_5G6Gevolution}.

%Edge intelligence can only reach its full potential when being combined with the appropriate AI algorithms.

It is expected that 6G networking systems will be in a massive scale supporting over 125 billion of connected devices by 2030. %Unfortunately, most of the data generated by these devices are random in nature and hand-labeling and extract useful information and features from these data is prohibitive.
%
%One of the key challenges
%The random nature of wireless networking dataset
%To address the final challenge, that is the lack of sufficient amounts of labeled dataset,
It is therefore critically important to develop an automatic data collecting, labeling, and processing architectural framework that allows the edge computing network to adapt and evolve by itself. %As mentioned earlier, there is no lack of data in future wireless system. However, collecting and labeling dataset in every possible conditions with  the wireless system can be affected by a large number of factors and hand-labeling a large number of
Self-learning is a emerging area in ML that allows agents to sense, learn, reason, decide, adapt, and evolve by themselves without any hand-labeling efforts nor human involvement\cite{Ye2017Selfsupervisedlearning,Jie2017CVPRSelfTaughtLearning,Real2020autoMLzero}. It leverages recent advances from a range of AI approaches including self-supervised learning\cite{Tran2019SelfsupervisedGAN}, self-taught learning\cite{Jie2017CVPRSelfTaughtLearning}, auto-ML\cite{Real2020autoMLzero}, etc., so as to achieve automatic label generation, feature extraction, representation learning, and model construction.
%both unsupervised learning % high data efficiency of the unsupervised learning approach with the high performance of the
%and supervised learning %. In particular, self-supervised does , but can
%by automatically creating labels from the raw dataset and use that to learn the representations in a supervised fashion.
It has achieved promising results under some specific use scenarios and has been considered as one of the most important future research directions of AI\cite{Real2020autoMLzero}. %Although it is still in the early stages of development, self-supervised learning-inspired architectures have already been recognized as one of the most promising solutions to fulfill 6G's vision of ubiquitous AI\cite{Li2019FLSurvey,Letaief2019Roadmapto6G}.

\begin{figure}
\centering
\includegraphics[width=3.5 in]{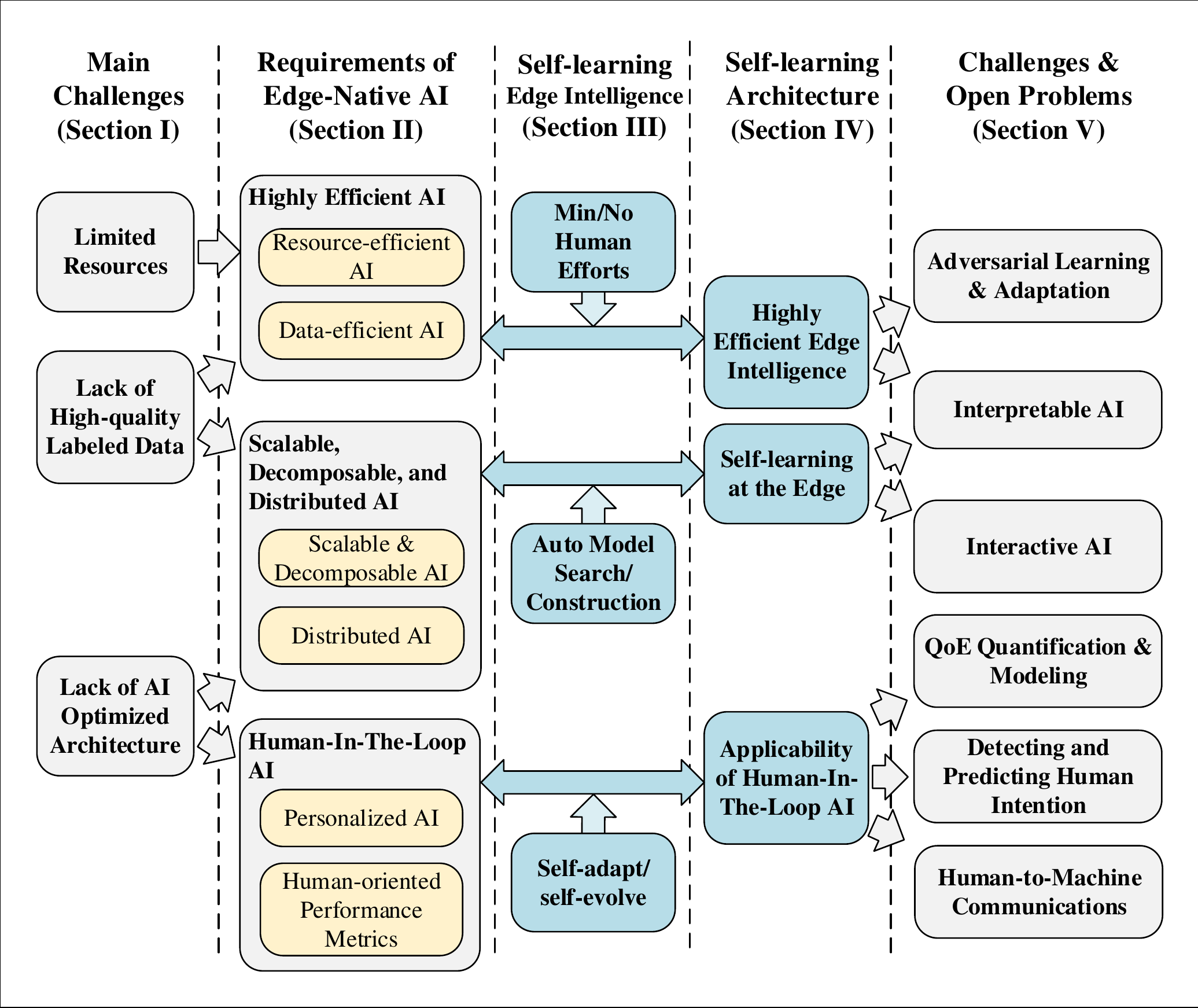}
\caption{Key requirements, potential self-learning-based solutions, and future challenges  of edge-native AI.}
\label{Figure_6GAIFLChallengeRelation}
%\vspace{-0.2in}
\end{figure}

The main contribution of this paper is to \blu{identify key requirements and trends that will drive edge intelligence for 6G, especially from the perspective of self-learning}. In particular, we propose a self-learning-based architecture and discuss its potential in addressing some of the key challenges in 6G. To the best of our knowledge, this is the first work that surveys self-learning and its possible applications in 6G. We summarize the main structure of this article including the key requirements, potential self-learning-based solutions, and future challenges to be discussed in the rest of this article in Fig. \ref{Figure_6GAIFLChallengeRelation}. %In particular, 6G needs to meet the following five requirements.
%
%
%
%The rest of this article is organized as follows. In Section \ref{Section_RequirementAI}, we describe the basic requirements and major challenges for edge-native AI in 6G, driven by recent trends in AI and edge intelligence. Basic concepts and requirements for self-learning edge intelligence are  discussed in Section \ref{Section_SelfLearningEdgeInte}. A self-learning-based architecture is proposed in Section \ref{Section_FL}. Finally, in Section \ref{Section_ChallengesandResearchTopics}, we describe the main challenges and important open research topics in self-learning-based 6G edge intelligence and conclude the article in Section \ref{Section_Conclusion}.

\begin{table*}[tbp]
\centering
\caption{Roles of AI in 4G, 5G, and 6G}
\vspace{-0.1in}
\label{Tabel_5G6GCompare}
\scriptsize
%\tiny
%\resizebox{\columnwidth}{!}{
\begin{tabular}{|l|l|l|l|}
\hline
%802.11ac
 & \makecell[c]{\bf Cloud-based AI (4G)} & \makecell[c]{\bf AI-enhance Functions (5G)} & \makecell[c]{ \bf Edge-native AI (6G)} \\
 \hline
{\bf Architecture} & Communication-oriented architecture & Service-based architecture (SBA) &
\makecell[l]{AI-native edge intelligence}  \\
\hline
\makecell[l]{\bf Functional\\ \bf Components} & \makecell[l]{Over-the-top AI applications \\ deployed in cloud data center \\ delivered via 4G networks} & \makecell[l]{Preset functional moduales\\ to monitor and enhance\\ performance of SBA} & \makecell[l]{Seamlessly integration of AI, \\communication network, and \\edge computing} \\
\hline
\makecell[l]{\bf Key Requirements} & \makecell[l]{Mostly applied in non-safety-related\\ applications} & \makecell[l]{Stringent latency and reliability\\ requirements in some use\\ scenarios, e.g., URLLC} & \makecell[l]{QoE guarantee with \\ self-adaptation \& self-learning\\ capability} \\
%\hline
%\makecell[l]{Performannce\\ Guarantee} & No QoS guarantee & \makecell[l]{10ms in URLLC use scenario} & \makecell[l]{10ms in URLLC use scenario} \\
%CSCC, TEC, ISHN, HOC, and Others%\footnote{Communication Services with Complex Constraints (CSCC), Time-engineered Communications (TEC), Integrated Services involving Heterogeneous Networks (ISHN), and Human-oriented Communication Service (HOS) are possible use scenarios for 6G which will be described in detail in Section \ref{Section_Requirements}.} \\
\hline
{\bf Applications} & \makecell[l]{Voice and image-recognition-based\\ virtual assistant app} & \makecell[l]{Self-driving vehicles, \\ smart factory, AR/VR} & \makecell[l]{Self-evolving smart city, interactive\\ holographic communication, highly\\ intelligent humanoid robot} \\
%Key New Technology & NS, FV, SDN, network Slicing mmWave, massive MIMO & Cross-disciplinary AI, distributed

%$3$ & $3$ & $15$ & $63$ & $8$ or $10$ msec   \\
%$4$ & $7$ & $15$ & $1023$ & $2$ or $10$ msec  \\
\hline
\end{tabular}
\vspace{-0.1in}
\end{table*}
\normalsize

%\begin{figure*}
%\centering
%\includegraphics[width=5.6 in]{Figure/5G6Gevolution.pdf}
%\caption{Key services, applications, and roadmap for 6G. }
%\vspace{-0.2in}
%\label{Figure_5G6Gevolution}
%\end{figure*}

\section{Requirements for Edge-native AI}
\label{Section_RequirementAI}

The substantial impact of 6G in terms of data collection, transportation, processing, learning, and service delivery will shape the evolution of network intelligentization, catalyzing the maturity of edge intelligence. In particular, next-generation edge-native AI technologies must be able to meet the following requirements raised by 6G.

\subsection{Highly Efficient AI}

\subsubsection{Resource-efficient AI}
Traditional wireless networks mainly focus on maximizing the data transportation capability of wireless resource such as spectrum and networking infrastructure.
%user-intended data.
However, with more computationally-intensive and data-driven AI tasks being adopted by 6G, the extra-resources required to perform AI-based process, including data coordination, model training, computing, caching, etc., must be carefully evaluated, quantified, and optimized. %Currently, there is still a lack of a general approach to quantify various resources required for most AI algorithms.  %For example, in commercial wireless networks, communication resources such as spectrum and network infrastructure are physically limited and expensive. Most of these resources must be preserved for transporting user data. In addition, exchanging a large amount of coordination and fused data for AI model training and learning throughout a large network will also result in intolerably high coordination and convergence delay that may significantly affect the speed and efficiency of decision making.
Although there are already some communication-efficient AI algorithms such as deep reinforcement learning, transfer learning, and federated learning that exhibit  reduced  communication overhead, these algorithms may still require considerable amount of resources compared to most data-centric applications\cite{XY2020WCSP}. Also, these algorithms can only be applied to some very specific learning tasks. %Their computing and communication efficiencies also vary significantly under different  network scenarios.

\subsubsection{Data-efficient AI}
As mentioned earlier, compared to computer vision systems, it is often more difficult to collect sufficient amounts of high-quality labeled dataset under each possible wireless environment and networking setup. Therefore, it is of critical importance to design data-efficient self-learning approaches that only require a limited or no hand-labeled data as input. There are already some data-efficient AI algorithms being introduced recently. For example, the self-supervised approach combines the advantages of both supervised and unsupervised AI approaches by automatically creating labels from the raw dataset for some pretext tasks and use that to learn the representations in a supervised fashion. Unfortunately, these approaches are still in their infant stage and can only be applied in a number of very limited tasks. % labels can be automatically created from the unlabeled raw dataset and then be fed into a supervised learning-based module to learn the requested representations.

 %Self-supervision is a learning framework in which a supervised signal for a pretext task is created automatically, in an effort to learn representations that are useful for solving real-world downstream tasks.

%\subsubsection{Fog/Edge-optimized AI} Most existing AI approaches focus on centralized data training and learning. With fog/edge computing becoming an indispensable part of 6G, there is a pressing need to develop fog/edge-optimized AI with data training, processing, and decision making spanning cloud data centers and multiple collaborative and/or distributed fog servers. There are two main challenges for developing fog/edge-optimized AI.

\subsection{Scalable, Decomposable, and Distributed AI}

\subsubsection{Scalable and Decomposable AI}
%The first one is to control and manage heterogeneous fog servers. More specifically, i
In contrast to a high-performance cloud data center, which is typically built based on the centralized architecture with a unified interface supported by compatible software and hardware components, edge computing is a highly distributed architecture, consisting of a large number of edge servers with heterogeneous processing and caching capabilities as well as energy and size constraints. Edge servers may also be deployed and managed by multiple service providers, supporting different software (Android, Ubuntu, or Windows) and hardware (ARM, RISC, or X86) platforms. It is therefore important to provide a scalable and decomposable data and task processing framework to allow parallel processing of tasks that span the cloud as well as multiple edge servers. One possible solution is to extend the network softwarization approach to edge hardware and software platforms. In this way, heterogenous hardware and software platforms can be abstracted into a set of virtual functions that execute  different AI tasks.

\subsubsection{Distributed AI}
One of the key challenges for edge intelligence is to design a simple,  scalable, and distributed AI approach that supports a large number of distributed edge servers as well as cloud data centers to jointly perform the same set of computational tasks. Distributed AI has attracted significant interest due to the recent popularity of federated learning and its extension-based solutions. However, both distributed AI and federated learning are still in their infancy. It is expected that the federated learning-enabled architecture will play an important role in the future evolution of distributed AI-based 6G services and applications. %We will give a more detailed discussion about that in Section \ref{Section_FL}.

\subsection{Human-In-The-Loop AI}
\subsubsection{Personalized AI} Personalized AI will play a key role in
6G to improve the decision making of AI algorithms and help
machines understand better about human users¡¯ preferences
and make more human-preferred decisions [12]. There are two
types of Human-In-The-Loop AI approaches. The first one is
to include human intelligence as part of the decision-making
process. For example, an AI algorithm can leverage human
intelligence to make decisions when the machine itself cannot
make correct decisions or the cost of making the incorrect
decision is high, e.g., in self-driving vehicular systems, each
vehicle can turn the control back to human drivers when it
faces unknown situations or cannot make a safety-guaranteed
driving decision. The other approach is to allow agents to
observe their past interactions with human users and learn to
improve the decision-making process.

\subsubsection{Human-oriented Performance Metrics} Instead of simply
focusing on maximizing traditional performance metrics,
such as throughput, network capacity, and convergence rate,
the performance of 6G and AI must be jointly measured
and evaluated by taking into consideration characteristics
and potential responses of users. In addition, with 6G and
mobile services becoming increasingly indispensable to human
society, it is also important to develop novel metrics that can
help evaluate the social and economic dimensions of 6G and
AI convergence.

\section{Self-learning Edge Intelligence}
\label{Section_SelfLearningEdgeInte}
Self-learning edge intelligence has %attracted significant interest recently due to its
the potential to significantly reduce the human efforts involved in data processing and model development by enabling self-detection and adaptation over unknown events, and most importantly, automatic model construction, learning, and evolving according to the changes of data features and environments. Self-learning edge intelligence in 6G must meet the following requirements.  %will be an essential part for 6G which is promised to have the following capacities.
\subsection{Minimized/No Human Effort}
%It is known that traditional AI solutions are highly tailored to specific tasks and dataset and each successful application may require a significant amount of efforts in data preprocessing and labeling as well as model construction and updating. In other words, t
The success of edge intelligence is expected to heavily rely on self-learning AI mechanisms with minimal human efforts for manual data processing and labeling. One promising solution is to leverage recent advances in self-supervised learning to enable model training based on automatically generated pseudo-labeled data, e.g., using self-supervised learning and generative neural networks such as generative adversarial networks (GANs) and variational autoencoders (VAEs). In the next section, we will introduce a self-supervised GAN-based architecture and present a case study to demonstrate the potential of automatic pseudo-labeling and data synthesizing in unknown service identification and classification.
\subsection{Automatic Model Search and Construction}
%Although a large number of ML algorithms and solutions being introduced during the past decade,
It is known that each individual ML model or algorithm possesses a specific structure and often comes with a set of sophisticated human expert-designed strategies or empirical rules for model construction and tuning. Recently, a new research direction, known as automatic ML (AutoML) has emerged. The goal of AutoML is to automate the architecture/model search across a range of existing ML approaches. It has attracted significant interest from both industry and academia. Developing simple and effective AutoML solutions will be essential for the practical implementation for self-learning edge intelligence networking systems.
%
% a promising research direction towards the self-learning edge intelligence.%
%
\subsection{Self-adaptation and Self-evolution}
Most existing AI approaches assume the system environment to be stationary within a certain period of time, and thus, decisions can be made based on a fixed known model or policy trained on a given dataset. For example, supervised learning, one of the mature and better-understood AI approaches, needs to be trained with properly-labeled data with pre-knowledge of all possible patterns. These approaches cannot work in continuously changing environments. Online learning and reinforcement learning-based approaches whose goal is to maximize the long-term reward have a strong potential to overcome the above challenges. For example, deep reinforcement learning has already been successfully adopted in Google's AlphaGo to beat the world's best human player in the game of Go. In addition to self-adapting to various known scenarios, self-learning edge intelligence should also have the capability to learn and self-evolve in unknown environments. Data synthesis is expected to play a major role in identifying and classifying various emerging unknown situations from a limited number of real-world data. As will be shown in the next section, compared to the traditional clustering solutions, novel data synthesizing solutions can significantly improve the accuracy and learning speed on recognizing and classifying unknown services.

% we will introduce a self-supervised GAN-based architecture and present a case study to demonstrate the potential of automatic pseudo-labeling and data synthesizing in unknown service discovery and classification.  %However, we are yet to see any  large-scale practical implementation of these learning-based approaches in communication systems.

%\subsection{Self-evolve}

%With new technologies being applied and integrated into future systems, networking systems can be dynamically changing and evolving into some future states that have not been encountered before.

 %Also, it is generally impossible for these approaches to offer any guaranteed performance under unknown events. %\cite{Stoica2017BerkeleyAI}.

%\subsubsection{Secure AI}
%With AI becoming more commonly used in network optimization and decision making, the privacy and security of the data as well as all the relevant data training and transporting processes become more important than ever. Both passive and active AI approaches need to be jointly considered and applied to further improve the security of AI-based network systems. For example, there are recent efforts to apply ``secure enclave," a closed-loop AI processing environment, to  shield AI-based applications from outside malicious attacks. In addition to executing  AI applications within secure enclaves, network operators can also inject known data probes into model training and using the observed output to monitor the data training process and detect whether the AI process has been compromised.

\section{A Self-learning Architecture for Edge Intelligence}
\label{Section_FL}

\begin{figure}
\centering
\includegraphics[width=3.3 in]{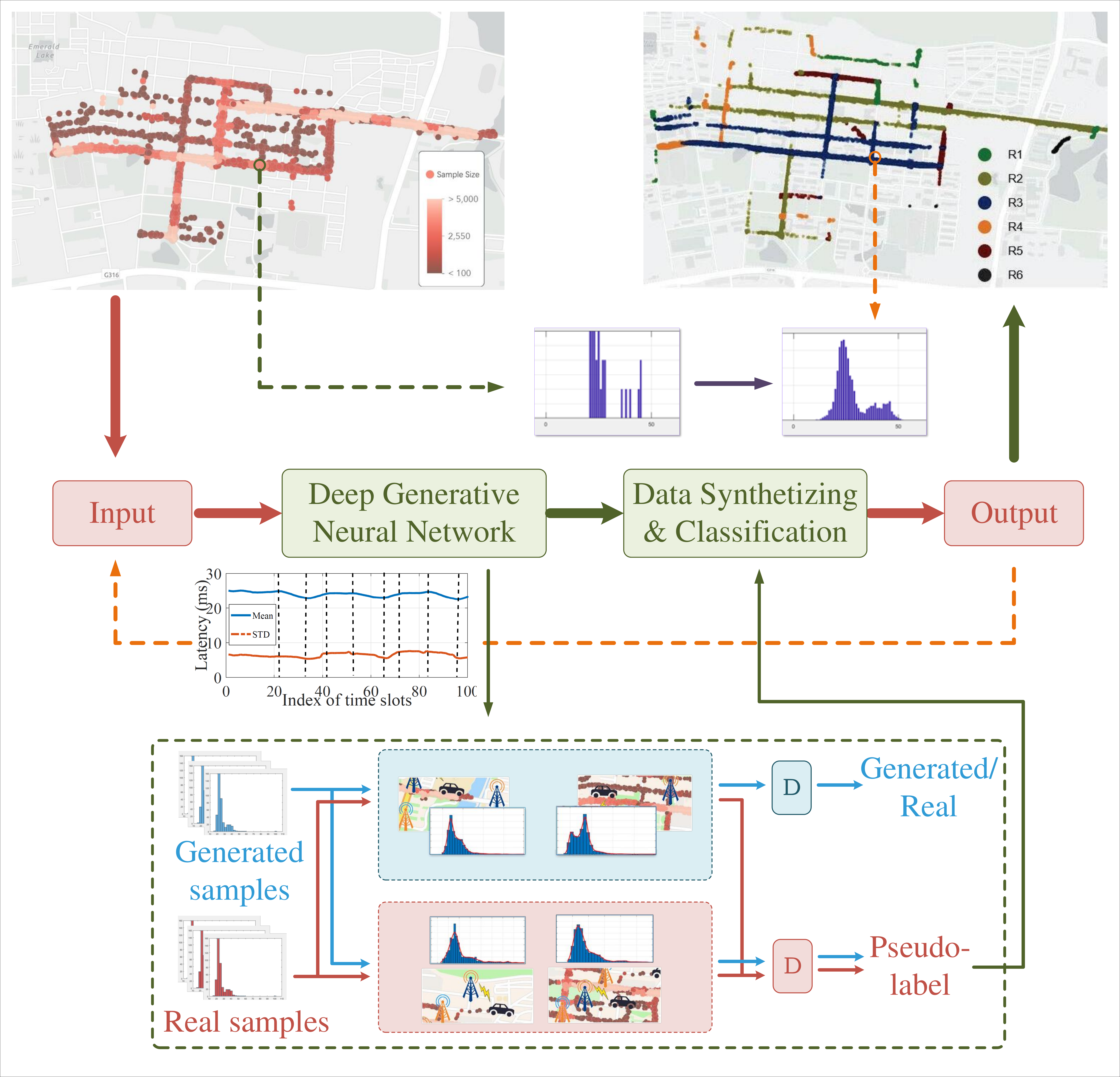}
\vspace{-0.1in}
\caption{The proposed self-learning architecture.}
\label{Figure_SL}
%\vspace{-0.2in}
\end{figure}

\begin{figure}
\centering
\includegraphics[width=3.3 in]{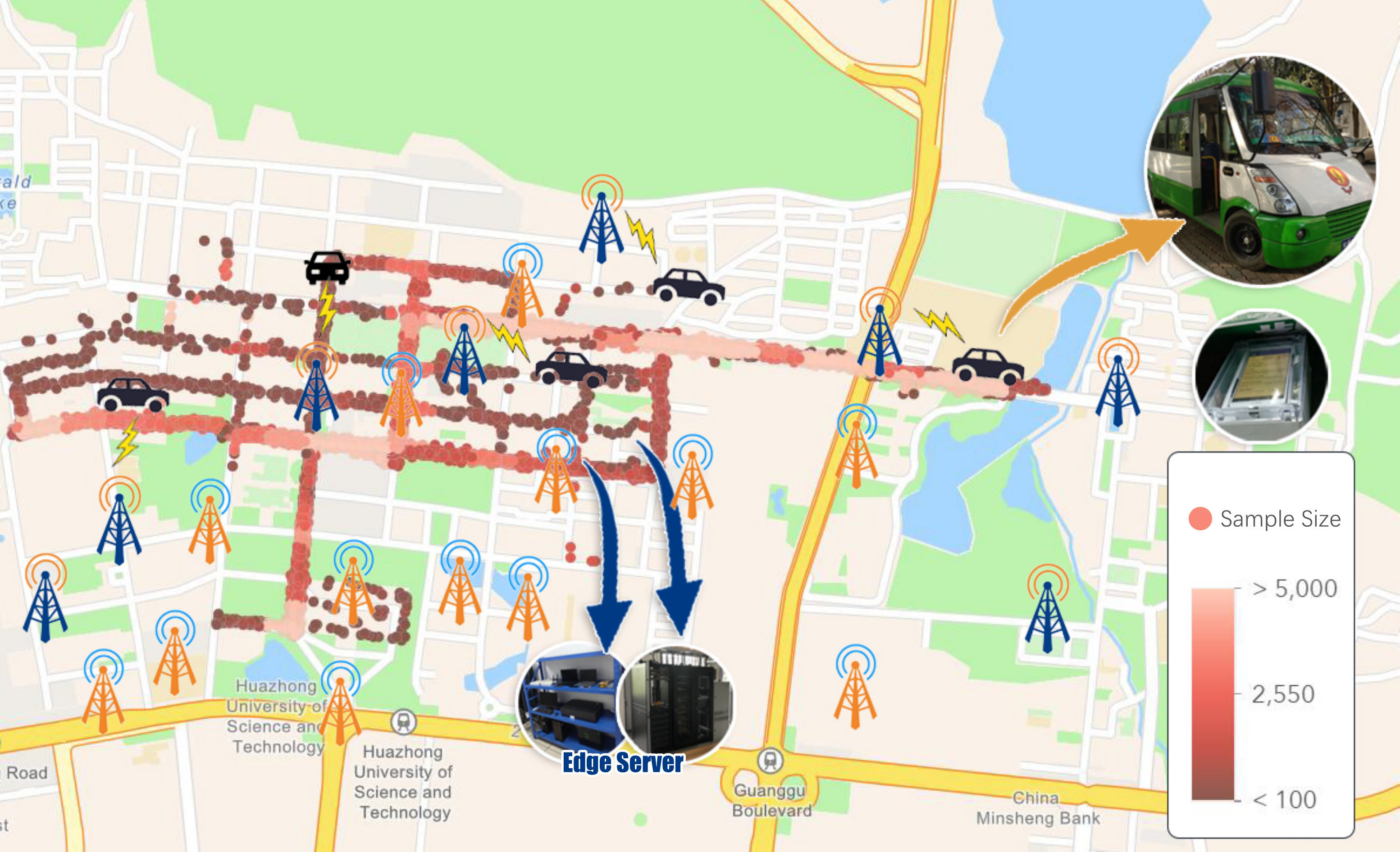}
\vspace{-0.1in}
\caption{A connected vehicular system for evaluating the performance of our proposed self-learning architecture.}
\label{Figure_HUSTCampus}
%\vspace{-0.2in}
\end{figure}

\begin{figure}[htbp]\label{fig equal_iid}
	\centering
		\begin{minipage}{0.5\linewidth}
			\centering
			\includegraphics[width=1.5 in]{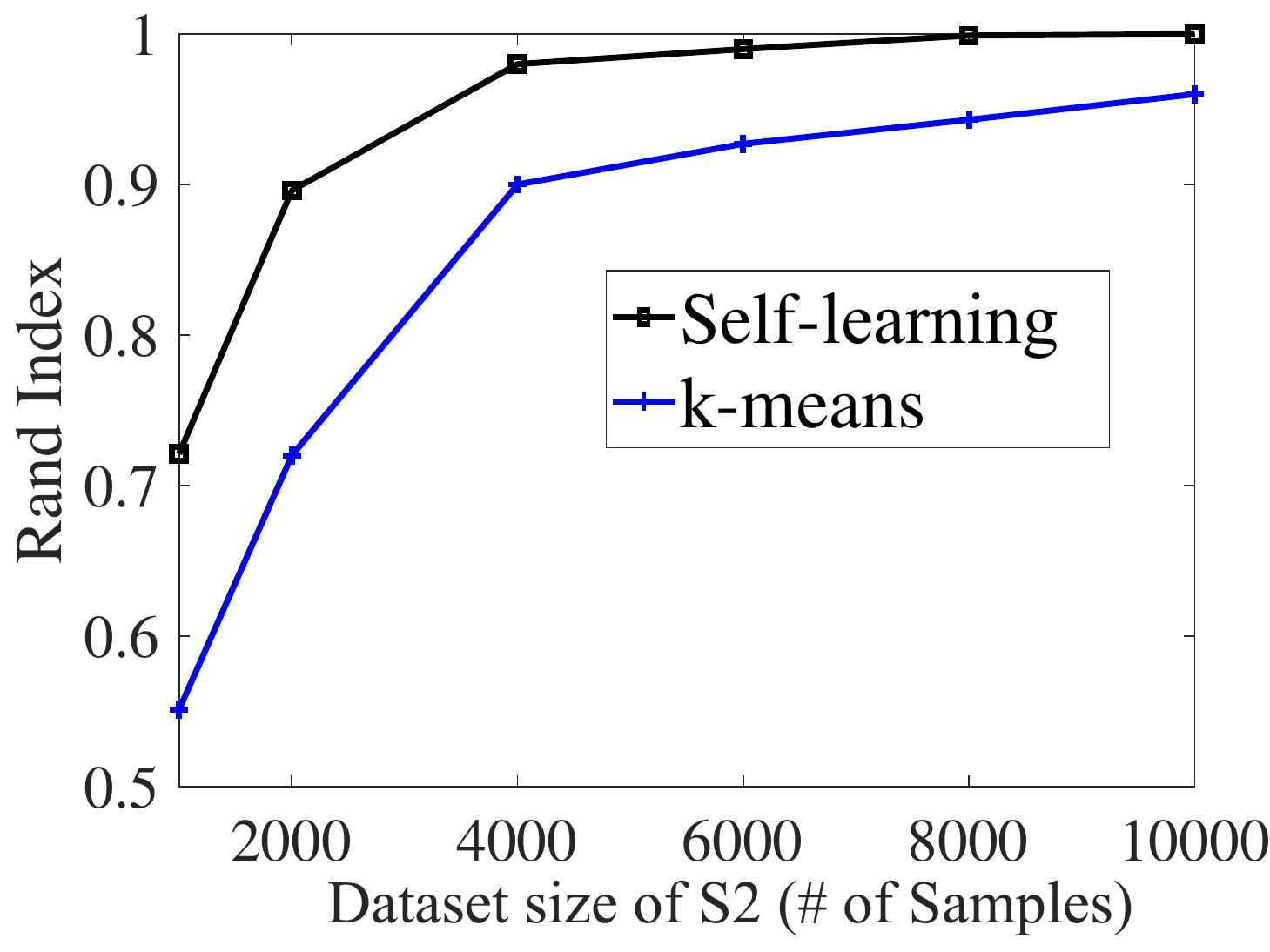}
\vspace{-0.1in}
			\caption{Classification performance\newline of the proposed architecture.}
\label{Figure_Classify}
		\end{minipage}%
		\begin{minipage}{0.5\linewidth}
			\centering
			\includegraphics[width=1.5 in]{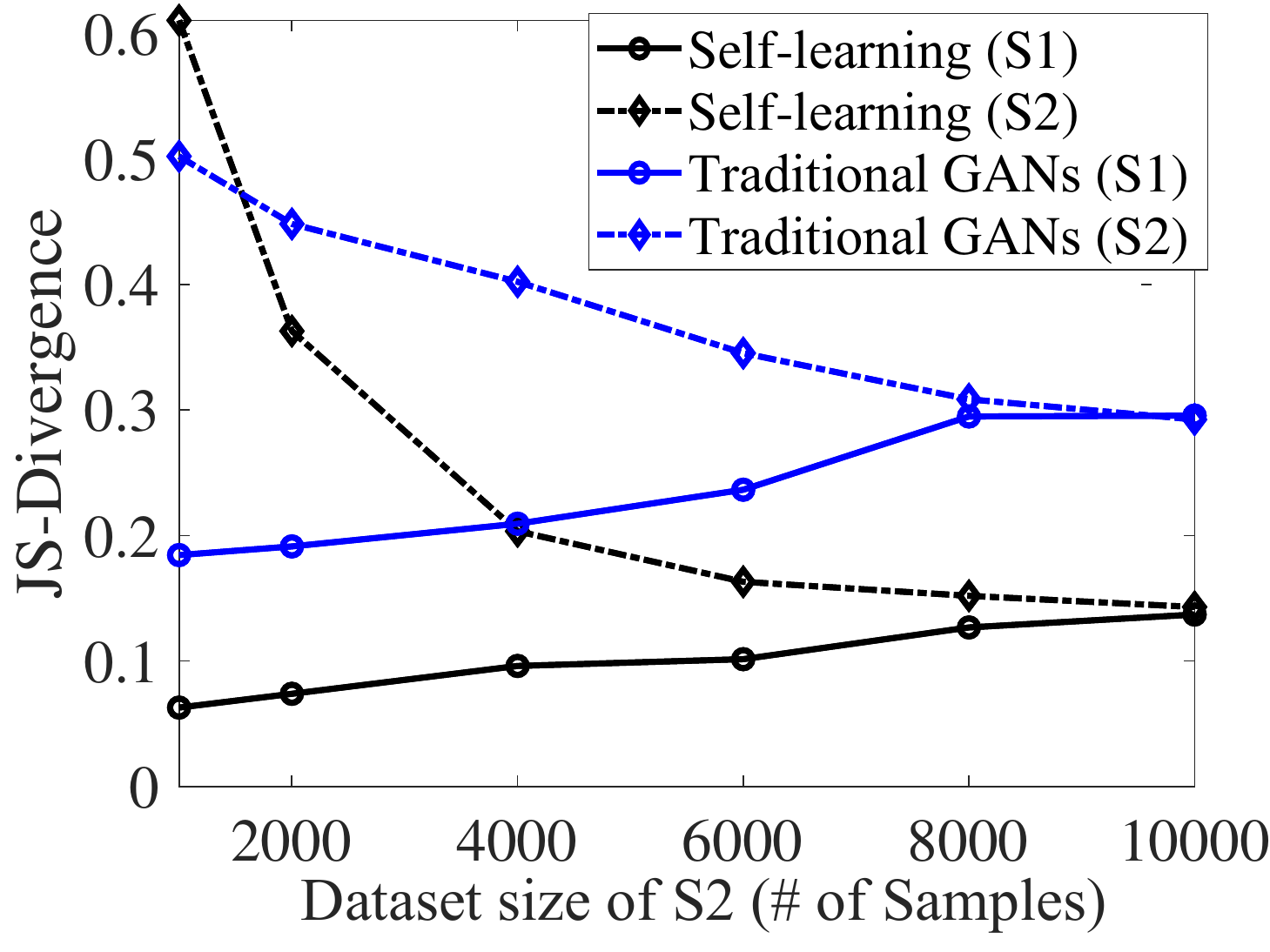}
\vspace{0.1in}
			\caption{JS-divergence between the synthetic data and real data for both services S1 and S2 under different dataset sizes of service S2.}
\label{Figure_Generate}
		\end{minipage}%
\end{figure}

The success of 6G will heavily rely on a simple and effective edge intelligence solution to meet all the above requirements and can, at the same time, self-adapt and self-evolve according to the future dynamics of the system. In this section, \blu{we introduce a simple self-learning architecture to demonstrate the potential improvement that can be achieved in unknown service traffic classification and prediction throughout a large-scale edge intelligence system. We use the connected vehicular network as a case study to evaluate the performance our proposed architecture under some specific scenarios. Finally, we discuss some possible extensions that can be brought by our self-learning architecture to meet some of the requirements of edge-native AI discussed in Section \ref{Section_RequirementAI}.}

\subsection{A Self-learning Architecture}
In this section, we propose a self-learning architecture based on self-supervised GAN with multiple generators\cite{Tran2019SelfsupervisedGAN} with the goal of automatically learning features and constructing ML models to identify and classify emerging unknown services  from raw crowdsourcing data distributed across a wide geographical area as illustrated in Fig. 2. Our proposed architecture exploits the generative learning feature of the GANs approach in which multiple generators are trained to produce synthetic data that can capture the mixture of distribution of traffic  data generated from multiple services across various locations in the coverage area. By introducing a classifier to maximize the distribution difference (e.g., measured by Kullback-Leibler (KL) Divergence) of synthetic data produced by different generators, we can prove that it is possible to train each generator to produce synthetic data samples that follow the same distribution as the real traffic data associated with each individual service. We can then leverage the self-supervised learning approach to automatically create pseudo-labels for the synthetic data produced by each generator. These created pseudo-labels will then be exploited to train a deep neural network model to identify and classify various service data across different locations throughout the service region\cite{Chen2019SelfsupervisedLearningRotations}. %The proposed approach proceeds as follows: (1) the real traffic data consisting of a mixture of multiple emerging unknown service traffic data is input to the model, (2) the discriminators and the generators will be jointly trained to produce a mixture of synthetic data that captures the distribution of the real traffic data, (3) a classifier is introduced to maximize the difference of distributions for synthetic data samples produced by different generators, (4) pseudo-labels will be created for synthetic data samples produced by each individual generator, and (5) the pseudo-labeled data will be input to the model in order to train a classifier that can identify and differentiate traffic data generated by various unknown services.

Different from the traditional GANs, the discriminators in the proposed architecture will perform two tasks at the same time: (i) recognizing whether the data is real or fake, and (ii) identifying which generator the data is associated with. The generator is unaware of these tasks but will learn to produce synthetic data samples with similar characteristics to the real recorded data. Both generator and discriminator try to optimize different and opposing objective functions until an Equilibrium solution is reached.

To reduce the computational load of each edge server, the generators and discriminators of the proposed architecture can be deployed separately across multiple edge servers, each trains the model with a subset of data. %In this case, the total computational capacity required for model training can be distributed across multiple low-cost computing servers and the computational power required for each individual edge server can be further reduced.
Another potential solution to reduce the computational complexity for training our GANs-inspired architecture is to leverage transfer learning and exploit the preliminary knowledge and model learned by other users or network components, so as to further reduce the computational load in model training.

\subsection{Case Study and Performance Evaluation}
We consider the latency-sensitive connected vehicular system consisting of six campus shuttles connected to two edge servers as well as a cloud data center via a 5G  network as a case study to evaluate the performance of our proposed architecture. We have developed a dedicated smart phone app\cite{XY2019AdaptiveFog} to monitor and keep track of the data delivery latency of 5G network connections between the moving shuttles and the associated edge servers as well as a cloud data center from a major service provider as shown in Fig. \ref{Figure_HUSTCampus}. In particular, we simulate two unknown connected vehicular services, labeled as S1 and S2, with different latency tolerances. Service S1 is the edge-based service that fully relies on the closest edge server to deliver the servers. Service S2 the edge/cloud-mixed service that equally divides its computational loads to be uploaded to the cloud and edge servers to perform computationally intensive services. We assume each vehicle can only observe the overall service latency data and cannot know whether the latency is caused by edge server or cloud data center. We apply our proposed self-learning architecture with one discriminator and two generators to classify the latency data associated with different services. We simulate the scenarios in which service S2 is an emerging service with different numbers of data samples being mixed with service S1.  %We consider training a self-learning architecture . %a workstation with an Intel(R) Core(TM) i9-9900K CPU$@$3.60GHz, 64.0 GB RAM @ 2133 MHz, 2 TB HD and two TITAN X (GP102). We simulate each edge server with a TITAN X GPU running on Ubuntu 16.04, python 3.6, cuda 10.0 and torch 1.3.1. The training process includes one discriminator and two generators, both of which are consisted of three MLPs.

To evaluate the service classification performance of our proposed architecture, in Fig. \ref{Figure_Classify}, we present the rand index (RI)%\footnote{RI is a commonly used metric for evaluating the clustering performance. %, defined as ${\rm RI} = {{\rm TP} + {\rm TN}\over {{\rm TP}+{\rm FP}+{\rm FN}+{\rm TN}}}$ where $\rm TP$ corresponds to the true positive decision assigning two similar data samples to the same cluster, $\rm TN$ is the true negative decision assigning two dissimilar data samples to different clusters. $\rm FP$ corresponds to the false positive decision assigning two dissimilar data samples to the same cluster, $\rm FN$ is the false negative decision assigning two similar data samples to different clusters.
%RI approaches 1 if the probability of false clustering decision is close to zero.}
of the clustering solutions of our architecture compared to the existing solutions such as k-means when the dataset sizes of service S2 vary. We can observe that the RI of our proposed architecture approaches 1 (probability of false clustering decision is close to zero) with only a limited number of service S2 data samples being available in the mixture of dataset consisting of both service traffics. To evaluate the quality of synthesized data samples generated by our architecture, we also compare the difference between the real data distribution and the distribution of synthesized data samples produced by our architecture measured by Jensen-Shannon (JS) divergence in Fig. \ref{Figure_Generate}. Our result shows that, by applying our proposed architecture, the distributions of synthesized data for both services approach those of the real service data. To summarize, our proposed architecture can classify unknown services from a complex mixture of service data without requiring any human labeled dataset. %In addition, the synthetic data samples produced by our architecture capture the distribution of real data associated with unknown services which can be applied to predict the future evolution of service-related data traffics as well as other prediction-based use cases such as network planning and anomaly detection.

%We compare the latency performance data learned and generated from our proposed architecture with that created by our hand-collected and labeled dataset. Our result shows that our proposed architecture is able to establish a sufficiently accurate statistic model of data delivery latency between connected vehicles and edge servers with less than one-hundredth of required data sizes.} %We also observe that although the latency performance also varies at different time, the temporal fluctuation of the latency performance exhibits periodic feature with .  % and the learned model
%To evaluate the performance of our proposed self-learning architecture, we use the latency performance dataset between a campus edge computer server connected via a 4G/5G mixed networks recorded over 4 months of measurements.
%The learned model as well as the knowledge related to the model can be transferred and utilized by other servers as well as the cloud data center for the benefit of others and/or coordination with other services and users across a wider geographic area\cite{Qiang2019FL}.

\subsection{Potential to Meet 6G Requirements}

\blu{The above self-learning architecture has the potential to be extended into more general forms to meet the various requirements raised by 6G.}

\subsubsection{Highly-efficient Edge Intelligence} In the above architecture, the main objective of the deep generative neural network is to produce synthetic data to capture the attributes  of real service data. % associated with unknown services.
In other words, it is unnecessary to collect a large number of high-quality manually-labeled data for each individual service across the entire coverage area. Also, since the synthetic data can be directly produced by the edge server, the data uploaded from the user as well as the total traffic transported throughout the network can be significantly reduced. % which further improves the utilization efficiency of the communication resources.
Our preliminary result also shows that the computational complexity of each edge server to perform the self-supervised GAN algorithm is also limited as long as the dataset sizes and heterogeneity among edge servers is limited, e.g., when %coverage area of
each edge server covers a smaller-sized area with a limited service demand. %is limited.

% Since the pseudo-labels is created and recorded automatically by the generator, the raw data collected in each individual location can  in each device need not to be uploaded to a centralized location, the total traffic transmitted on the network will be reduced. Recent works show that the communication overhead of FL can be further reduced by compressing the data size of each device-side updating or reducing the model update frequency, e.g., by avoiding updates from devices with limited contributions to model convergence\cite{Li2019FLSurvey}.

\subsubsection{Self-adaptation and self-evolution at the Edge}
%
%\subsubsection{Heterogeneous Data Support}
Datasets collected by different devices can vary significantly in terms of their statistical features due to their different service types, use scenarios, user preferences, etc. The self-learning architecture is able to produce synthetic data that captures the distribution of any type of raw data input and can be automatically adapted to the change of the data types as well as other feature dynamics. Also the applicable areas and achievable solutions of self-learning architecture can be further enriched by recent developments on the integration of other state-of-the-art AI solutions, such as federated learning, semi-supervised learning\cite{Ye2017Selfsupervisedlearning}, reinforcement learning, %\cite{liu2019FLreinforcementlearning},
 transfer learning, and autoML.   %model training among devices with non-identically distributed (non-IID) datasets\cite{Bonawitz2019FLMassiveDeploy}.

%\subsubsection{FL for Privacy Preservation}
%The updated information uploaded by each device is inspectable but yet cannot be used to recover any useful information about the local dataset. The security level of FL can be further improved by adopting more advanced encryption and security measures, such as  secure aggregation at each fog server and user-level differential privacy\cite{Li2019FLSurveyPrivacy}.

%\subsubsection{Massive-scale Deployment Support} Since each edge server is only focusing on producing synthetic data that mimic the features of the locally recorded raw dataset, the proposed self-learning architecture can be directly applied into a massive-scale network deployment. %\cite{McMahan2017FLfirstpaper}. %In particular, experimental results show that FL can converge to the optimal solution even when the number of devices participating in model training is much larger than the average number of samples in the dataset of each device\cite{Bonawitz2019FLMassiveDeploy}.

\subsubsection{Applicability of Human-In-The-Loop AI} The prior knowledge and service preference of human users can be exploited to further improve the efficiency of the above architecture. In particular, human knowledge or any prior information can be directly used to design more pretext tasks for the self-supervised learning approach to further improve the self-learning performance. Also, the architecture also allows agents or system components to interact with human users by self-adapting to the changes of the environment or networks caused by human usage.  % if the predicted performance of the wireless connectivity between users and edge severs in some locations is estimated to be  almost impossible to meet the stringent requirements of requested services, e.g., safety-related services. it  can initiate a warning signal to all the connected vehicles heading to these locations and return the control back to human users/drivers.

%\subsubsection{Flexible Architecture}  %\cite{Augenstein2019FLGAN}.

\section{Challenges and Open Research Topics}
\label{Section_ChallengesandResearchTopics}

\subsection{Adversarial Learning and Adaptation}
AI-enabled 6G will be exposed to various novel attacks that aim to compromise the data training and decision making process. It is, therefore, important to develop effective self-adaptive methods to learn, detect, and defend against these attacks. Currently, there is no effective solution to protect against many data-related attacks, such as data evasion and poisoning attacks. One possible solution is to include the data affected by these attacked as the input and build a self-learning system that is resilient to various types of attacks. For example, if the impact of these attacks on model training and data processing can be carefully evaluated, network providers can leverage some existing approaches such as replay-with-simulating to efficiently eliminate the adverse effect caused by these attacks on the learned model.

%\subsection{Explainable AI}

\subsection{Interpretable AI}
Wireless systems need to be engineered with justifiable results and performance guarantees when being integrated with different network components and for different requirements. Unfortunately, most existing AI solutions, especially deep learning-based approaches, follow a black-box approach without any clear explanation about why and how the approach led to the given outcome. Developing explainable AI with interpretable and predictable outcomes is one of the key challenges for applying AI in 6G systems.

\subsection{Quality-of-Experience (QoE) Quantification and Modeling} 6G is expected to  focus more on optimizing and improving users' QoE instead of the QoS. QoE is more closely
related to the subjective experience of users. It not only depends on the hardware and software configuration and capacity, but can also be influenced by a wide range of human-related factors such as personal feelings, emotions, and past experiences. It is also affected by age, gender, personality of the users, as well as some environmental conditions such as service time, location, and physical environmental elements. There is still lacking a general and effective  model formulation that can quantify the QoE of human users considering all the above elements and conditions.
\subsection{Interactive AI} With the popularity of AI-enabled smart
devices and network elements, interactions between networking
elements and mobile devices are expected to be much more
complicated than before. The service performance in this case
will not only be affected by the hardware and software capacity
of each device but also its level of intelligence, including the
response mechanism and learning speed and capability of all
interacting users as well as their past and current interactions.

\subsection{Detecting and Predicting Human Intention}
Due to the random nature of human beings, user traffic and demands exhibit temporal and spatial fluctuations. For example, an autonomously driven vehicle can frequently switch back and forth between manual mode (with human control) and self-driving mode, causing large fluctuations in driving-assistant-related data traffic.
This will affect the stability and robustness of network systems, especially in densely deployed networks. Developing an AI-based solution that can detect and keep track of human intention as well as the driving mode of vehicles will help the network to be better prepared with improved reliability and robustness. It will also help the network to understand more the human users' real-time QoE and adjust the service performance accordingly.

\subsection{Intelligent Human-to-Machine Communications}
%With the popularity of AI-enabled smart devices and network elements, the interaction between networking elements and human users are expected to be much more complicated than ever. The service performance in this case will not only be affected by the hardware and software capacity of each device but also its level of intelligence, capability of sensing human-intention and making human-like decision, as well as the semantics of interaction, including the interacting users' real intentions (goal of communications), response mechanism, learning speed, capability of understanding each other's meanings as well as their interacting history. 
It is expected that 6G will be supporting a much wider range of novel interactive services and applications involving direct or indirect human-to-machine communications. A universal and human-oriented networking framework in which different components of networking systems can sense, communicate, and interact according to the real intentions of human users, irrespective of different interfaces, backgrounds, languages, and protocols, is expected to emerge in 6G era. % based on the meaning of the message. In this framework, users and devices with different interfaces, backgrounds, languages, and protocols will be able to communicate and interact with each other in a highly efficient way.

\section{Conclusion}
\label{Section_Conclusion}

This article provided an overview of a possible research roadmap for 6G edge intelligence, from the perspective of self-learning AI. Potential requirements and challenges of edge-native AI in 6G have been identified. Motivated by the major challenges for incorporating AI in wireless networks, that include resource limitation, lack of labeled data, and no AI-optimized architecture, we proposed a self-learning architecture that supports automatic data learning and synthesizing at the edge of the network. \blu{We evaluate the performance of our proposed self-learning architecture in a campus shuttle systems connected to edge servers via a 5G network. Our result suggests that our proposed architecture has the potential to further improve the data classification and synthesizing performance even for unknown services in the edge computing network under certain scenarios. The potential of self-learning AI to address some other novel challenging issues of 6G edge intelligence is also discussed.} %Although self-learning edge intelligence has a great potential, many fundamental technologies are still in the early developments.
We hope this article will spark further interest and open new research directions into the evolution of self-learning and its applications towards edge intelligence  in 6G.

\section*{Acknowledgment}
%We would like to thank Prof. Ross Murch at the Hong Kong University of Science and Technology and Prof. Luiz DaSilva at the CONNECT, Trinity College Dublin, for their helpful comments.
This work was supported in part by the National Natural Science Foundation of China under Grants 62071193 and 61632019, the Key R \& D Program of Hubei Province of China under Grant 2020BAA002, China Postdoctoral Science Foundation under Grant 2020M672357, and U.S. National Science Foundation under Grants CCF-1908308 and CNS-1909372.

\bibliography{DeepLearningRef}
\bibliographystyle{IEEEtran}

%%\newpage
\begin{IEEEbiographynophoto}{Yong Xiao}(S'09-M'13-SM'15) is a professor in the School of Electronic Information and Communications at the Huazhong University of Science and Technology (HUST), Wuhan, China. He is also the associate group leader in the network intelligence group of IMT-2030 (6G promoting group).
His research interests include machine learning, game theory, and their applications in cloud/fog/mobile edge computing, green communication systems, wireless networks, and Internet-of-Things (IoT).
\end{IEEEbiographynophoto}

\begin{IEEEbiographynophoto}{Guangming Shi} (SM'06) received the M.S. degree in computer control, and the Ph.D. degree in electronic information technology from Xidian University, Xi¡¯an, China, in 1988, and 2002, respectively. His research interest includes artificial intelligence, semantic analysis, brain-inspired computing, and signal processing. He is a Professor with the School of Artificial Intelligence, Xidian University. He is the chair of IEEE CASS Xi'an Chapter, senior member of ACM and CCF, Fellow of Chinese Institute of Electronics, and Fellow of IET. He was awarded Cheung Kong scholar Chair Professor by the ministry of education in 2012. And he won the second prize of the National Natural Science Award in 2017.
%is a professor in the School of Artificial Intelligence, the Xidian University. He is also the Vice President of Xidian University. %He has authored and co-authored more than 200 research papers. His current research interests include compressed sensing, and implementation of algorithms for intelligent signal processing. %He has authored or co-authored more than 200 research papers.
%His research interests include compressed sensing, intelligent signal processing, deep neural networks, and artificial intelligence.
\end{IEEEbiographynophoto}

\begin{IEEEbiographynophoto}{Yingyu Li} is a postdoc researcher in the School of Electronic Information and Communications at the Huazhong University of Science and Technology (HUST), Wuhan, China. Her research interests include IoT, networking data analysis.
\end{IEEEbiographynophoto}

\begin{IEEEbiographynophoto}{Walid Saad} (S'07-M'10-SM'15-F'19) is a Professor at the Department of Electrical and Computer Engineering at Virginia Tech. %, where he leads the Network Science, Wireless, and Security (NetSciWiS) laboratory, within the Wireless@VT research group.
His  research interests include wireless networks, machine learning, game theory, cybersecurity, unmanned aerial vehicles, and cyber-physical systems.
\end{IEEEbiographynophoto}

\begin{IEEEbiographynophoto}{H. Vincent Poor} (F'87) is the Michael Henry Strater University Professor of Electrical Engineering at Princeton University. His interests include information theory, machine learning and networks science, and their applications in wireless networks, energy systems, and related fields. Dr. Poor is a Member of the National Academy of Engineering and the National Academy of Sciences, and a Foreign Member of the Chinese Academy of Sciences and the Royal Society. He received the Marconi and Armstrong Awards of the IEEE Communications Society in 2007 and 2009, respectively, and the IEEE Alexander Graham Bell Medal in 2017.
\end{IEEEbiographynophoto}

% that's all folks
\end{document}